\documentclass[structabstract]{aa}  
\usepackage{amssymb}
\usepackage[usenames, dvipsnames]{color}
\usepackage{graphicx}
\usepackage{natbib}
\usepackage{txfonts}
\usepackage{units}
\usepackage{url}
\newcommand\tsp{\mbox{$\;\!$}}
\graphicspath{{//home/meetu/article2/manuscript/}}
\usepackage{hyperref}
\hypersetup{
    final=true,
    pageanchor=true,
    colorlinks=true,
    breaklinks=true,
    linkcolor=blue,
    citecolor=blue,
    urlcolor=blue,
    pdfpagemode=UseNone,
    pdftitle={Horizontal flow fields observed in Hinode G-band images.
        III. The decay of a satellite sunspot in active region NOAA~10930 and
             the role of magnetic flux removal in flaring},
    pdfauthor={M. Verma and C. Denker},
    pdfsubject={Solar Physics},
    pdfkeywords={Sun: chromosphere, Sun: photosphere, Sun: surface magnetism,
    Sun: sunspots, Techniques: image processing, Methods: data analysis}}
\newcommand\phn{\phantom{0}}

\newcommand\arcdeg{\mbox{$^\circ$}}

\begin{document}

%###############################################################################
%#
%#    TITLE
%#
%###############################################################################

\title{Horizontal flow fields observed in\\
       Hinode G-band images}
\subtitle{III. The decay of a satellite sunspot\\
               and the role of magnetic flux removal in flaring}

\author{M.\ Verma \and C.\ Denker}

\institute{Leibniz-Institut f\"ur Astrophysik Potsdam (AIP),
    An der Sternwarte 16,
    D-14482 Potsdam,
    Germany\\
    \email{mverma@aip.de, cdenker@aip.de}}

\date{Received 28 May 2012; accepted 24 July 2012}

% \abstract{}{}{}{}{}
% 5 {} token are mandatory
\abstract
% context heading (optional)
{Emergence of magnetic flux plays an important role in the initiation of flares.
However, the role of submerging magnetic flux in prompting flares is more
ambiguous, not the least because of the scarcity of observations.}
% aims heading (mandatory)
{The flare-prolific active region NOAA~10930 offered both a developing 
$\delta$-spot and a decaying satellite sunspot of opposite polarity. The
objective of this study is to characterize the photometric decay of the
satellite sunspot and the evolution of photospheric and chromospheric horizontal
proper motions in its surroundings.}
% methods heading (mandatory)
{We apply the local correlation tracking technique to a 16-hour time-series of
\textit{Hinode} G-band and Ca\,\textsc{ii}\,H images and study the horizontal
proper motions in the vicinity of the satellite sunspot on 2006
December~7. Decorrelation times were computed to measure the lifetime of solar
features in intensity and flow maps.}
% results heading (mandatory)
{We observed shear flows in the dominant umbral cores of the satellite sunspot.
These flows vanished once the penumbra had disappeared. This slow penumbral
decay had an average rate of 152~Mm$^2$~day$^{-1}$ over an 11-hour period.
Typical lifetimes of intensity features derived from an autocorrelation analysis
are 3--5~min for granulation, 25--35~min for G-band bright points, and up to
200--235~min for penumbrae, umbrae, and pores. Long-lived intensity features
(i.e., the dominant umbral cores) are not related to long-lived flow features in
the northern part of the sunspot, where flux removal, slowly decaying
penumbrae, and persistent horizontal flows of up to 1~km~s$^{-1}$ contribute to
the erosion of the sunspot. Finally, the restructuring of magnetic field
topology was responsible for a homologous M2.0 flare, which shared many
characteristics with an X6.5 flare on the previous day.}
% conclusions heading (optional)
{Notwithstanding the prominent role of $\delta$-spots in flaring, we conclude
based on the decomposition of the satellite sunspot, the evolution of the
surrounding flow fields, and the timing of the M2.0 flare that the vanishing
magnetic flux in the decaying satellite sunspot played an instrumental role in
triggering the homologous M2.0 flare and the eruption of a small H$\alpha$
filament. The strong magnetic field gradients of the neighboring $\delta$-spot
merely provided the vehicle for the strongest flare emission about 10~min after
the onset of the flare.}

\maketitle

%###############################################################################
%#
%#    INTRODUCTION
%#
%###############################################################################

\section{Introduction}

A theoretical description of flares has to explain their underlying cause and
the origin of the various observed features (e.g., flare ribbons, loops,
emissions across all wavelength regimes, and particle acceleration).
\citet{Priest2002} review recent flare models--starting with some type of
magnetohydrodynamic (MHD) catastrophe followed by magnetic reconnection--with an
emphasis on three-dimensional models, which are required for a full appreciation
of the dynamics in complex magnetic field topologies. Beyond modeling efforts, a
wealth of new data became available by space missions such as RHESSI, Yohkoh,
TRACE, and SoHO. Observations in the extreme ultra-violet (EUV) and in soft/hard
X-rays revealed a plethora of transition region and coronal arcades and loops as
summarized by \citet{Benz2008}, who discusses energetic and eruptive events as
well as the nature of energy release and particle deposition. More recently, the
relevant MHD processes such as flux emergence, formation of a current sheet,
rapid dissipation of electric current, shock heating, mass ejection, and
particle acceleration have been recounted by \citet{Shibata2011}. 

% FIGURE 1
\begin{figure*}
\includegraphics[width=\textwidth]{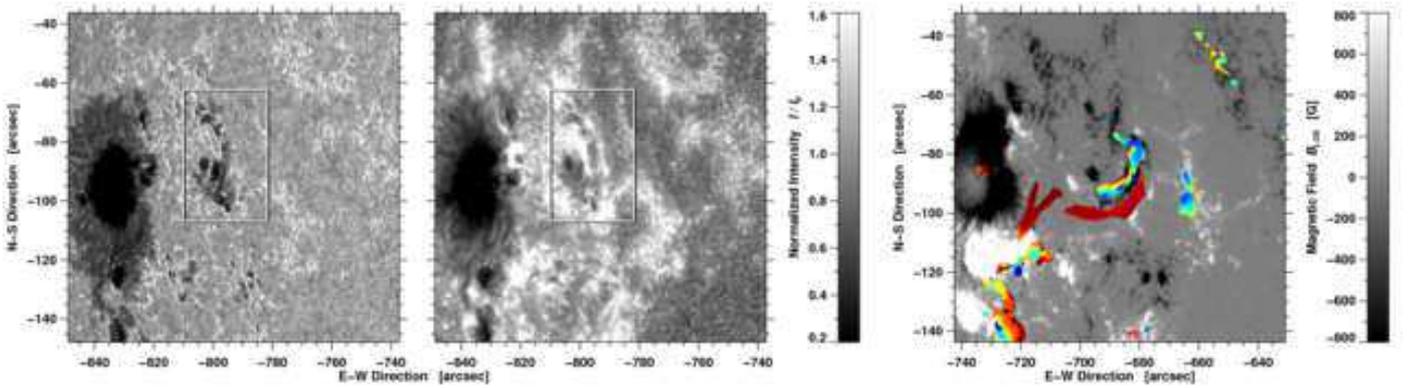}
\caption{First calibrated G-band (\textit{left}\tsp) and Ca\,{\sc ii}\,H
    (\textit{middle}\tsp) image in a 16-hour time-series of active region
    NOAA~10930 at 2:30~UT on 2006 December~7. The region in the white rectangle
    is the region-of-interest, i.e., a small satellite sunspot. SOT/NFI
    magnetogram (\textit{right}\tsp) captured at 18:46~UT on 2006 December~7.
    The rainbow-colored regions are the flare kernels derived from 32 Ca\,{\sc
    ii}\,H images covering the M2.0 flare for 16~min from 18:47~UT to 19:03~UT.
    The time progresses from blue to red. A Big Bear Solar Observatory full-disk
    H$\alpha$ \citep{Denker1999a} image taken at 18:39~UT is used to trace a
    small filament (reddish
    colors), which consists of a \textsf{V}- and sickle-shaped region. The
    annotation of the axes refers to heliocentric coordinates given in seconds
    of arc.}
\label{FIG01}
\end{figure*}

Newly emerging flux has been linked to solar flares, whereas flares related to
sunspot decay are not broadly discussed in literature. Simultaneous emergence
and submergence of magnetic flux has been explored by \citet{Kalman2001} for the
two (recurring) active regions NOAA~6850 (6891) and 7220 (7222). Based on X-ray
observations no direct interaction of new and old magnetic flux was evident.
However, flaring was observed for active region NOAA~6891 where only one
magnetic polarity submerged. The magnetic field evolution was comparatively slow
in these cases with time scales ranging from several hours to days. An
intermediate case has been discussed by \citet{Wang2002b}, who observed an M2.4
flare associated with rapid penumbral decay (within a time period of just a few
minutes) immediately after the flare, which was followed by the slow decay
(three hours) of the remaining umbral core. The first phase of sunspot decay,
i.e., rapid penumbral decay, has been established as an important signature of
flare-related photospheric magnetic field changes \citep[see e.g.,][]{Wang2004,
Deng2005, Liu2005, Sudol2005, Ravindra2012}. In our study of the flaring active
region NOAA~10930, we focus on the gradual decay of a satellite sunspot and
discuss its relationship to a developing $\delta$-spot of opposite polarity.

Active region NOAA 10930 was the first complex sunspot group producing X-class
flares that was observed by the Japanese \textit{Hinode} mission. The region was
the source of the well-studied X3.4 solar flare on 2006 December~13
\citep[e.g.,][]{Schrijver2008}. For this highly active time period
\citet{Tan2009} applied local correlation tracking (LCT) to study horizontal
proper motions related to
penumbral filaments in a rapidly rotating $\delta$-spot \citep{Min2009}. Both
studies found twisted penumbral filaments, shear flows, sunspot rotation, and
emerging flux at various locations within the active region. However, for the
time just after the region rotated onto the solar disk, fewer studies were
published, which mostly focused on the X6.5 flare on 2006
December~6 \citep[e.g.,][]{Wang2012}.

\citet{Bala2010} described a prominent Moreton wave having an angular extent of
almost 270\arcdeg, which was initiated by the X6.5 flare. The radiant point of
the Moreton wave appeared to be located at a small satellite sunspot to the west
of the major sunspot, whereas X-ray, white-light, and G-band emissions were
centered on the developing $\delta$-spot to the south. The strongest changes of
the magnetic force also occurred at this location \citep[cf.,][]{Fisher2012}.
Rapid penumbral decay and changes in the horizontal flow fields associated with
the X6.5 flare were discussed by \citet{Deng2011}. They observed the enhanced
and sheared Evershed flow along the magnetic neutral line separating the main
and $\delta$-spots. They concluded that the increased flow speed is not
associated with new flux emergence in the active region. In the present study,
we follow up the evolution of active region NOAA~10930 after the X6.5 flare
leading up to a homologous M2.0 flare on 2006 December~7.

%###############################################################################
%#
%#    OBSERVATION
%#
%###############################################################################

\section{Observations}

% TABLE 1
\begin{table}
\caption{Flare list for 2006 December~7 with starting, peak, and end times as
well as flare sites in heliocentric coordinates and X-ray flare
class.}\label{TAB01}
\tiny
\begin{tabular}{cccccc}
\hline\hline
Start Time & Peak Time & End Time & \multicolumn{2}{c}{Position} & X-Ray Class\rule[-2mm]{0mm}{5mm}\\
\hline
03:32~UT & 03:36~UT & 03:39~UT & E766\arcsec & \phn S86\arcsec  & C2.0\rule[-1mm]{0mm}{4mm}\\
04:27~UT & 04:45~UT & 05:09~UT & E764\arcsec & S120\arcsec & C6.1\rule[-1mm]{0mm}{4mm}\\
10:49~UT & 11:48~UT & 12:57~UT & E720\arcsec & S120\arcsec & C1.1\rule[-1mm]{0mm}{4mm}\\
14:49~UT & 15:15~UT & 15:33~UT & E709\arcsec & S120\arcsec & C1.2\rule[-1mm]{0mm}{4mm}\\   
18:20~UT & 19:13~UT & 19:33~UT & E687\arcsec & S103\arcsec & M2.0\rule[-1mm]{0mm}{4mm}\\
\hline
\end{tabular}\\
\tiny \hspace*{5mm}
\parbox{80mm}{\vspace*{-2mm} 
\begin{itemize}
\item[Note:] Figure~\ref{FIG02} provides a graphical representation of
flare timing and the GOES X-ray flux. Data were provided by NOAA's National
Geophysical Data Center (NGDC). 
\end{itemize}}
\end{table}

Active region NOAA~10930 appeared on the eastern solar limb on 2006 December~5.
The sunspot group was classified as a $\beta \gamma \delta$-region exhibiting a
complex magnetic field topology, and it produced numerous C-, M-, and X-class
flares. On 2006 December~7, we have analyzed 16~hours of data during the time
period from 02:30~UT to 18:30~UT about eight hours after the X6.5
flare. The region was located at heliocentric coordinates E800\arcsec\ and
S85\arcsec\ ($\mu = 0.56$, where $\mu = \cos(\theta)$ is the cosine of the
heliocentric angle $\theta$). The sequence with a cadence of 30~s is comprised
of 1920 G-band and Ca\,\textsc{ii}\,H images (left and middle panels of
Fig.~\ref{FIG01}) captured by the Solar Optical Telescope
\citep[SOT,][]{Tsuneta2008} on board the Japanese \textit{Hinode} mission
\citep{Kosugi2007}. We dropped every second image from the time-series because a
cadence of 60~s is sufficient for measuring horizontal proper motions.
A SOT/NFI magnetogram (right panel of
Fig.~\ref{FIG01}) serves as reference to illustrate the magnetic configuration
of NOAA~10930, which is later used in the discussion
of the M2.0 flare at 18:47~UT (see Sect.~\ref{SEC03.5}). The G-band and
Ca\,\textsc{ii}\,H images are $2 \times 2$-pixel binned with an image scale of
0.11\arcsec\ pixel$^{-1}$. Thus, the $1024 \times 1024$-pixel images have a FOV
of $111\arcsec \times 111\arcsec$. After basic data calibration, the images were
corrected for geometrical foreshortening and resampled onto a regular grid of
80~km $\times$ 80~km \citep[see][]{Verma2011}.

% FIGURE 3
\begin{figure*}
\includegraphics[width=\textwidth]{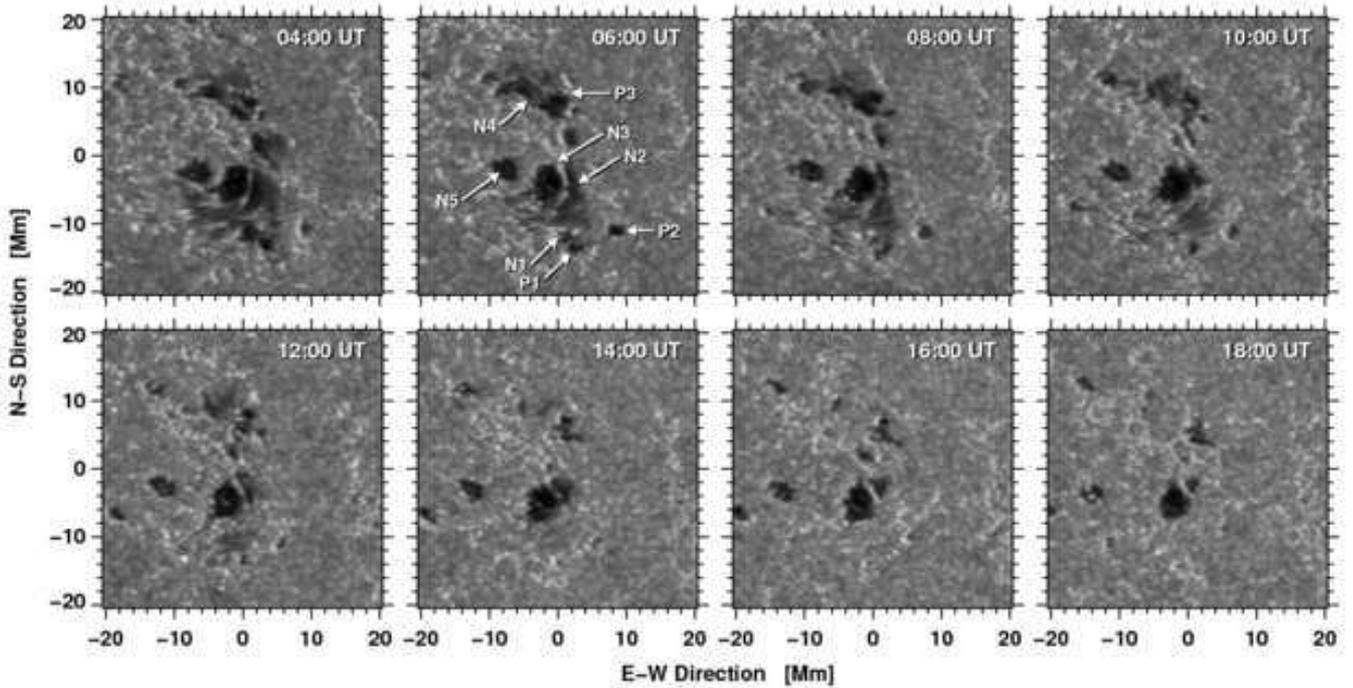}
\caption{Time-series of calibrated and geometrically corrected G-band images of
    the satellite sunspot in active region NOAA~10930 depicting various stages
    of its decay. The labels \textsf{N} and \textsf{P} refer to magnetic
    features with negative and positive polarities, respectively.}
\label{FIG03}
\end{figure*}

\begin{figure}
\includegraphics[width=\columnwidth]{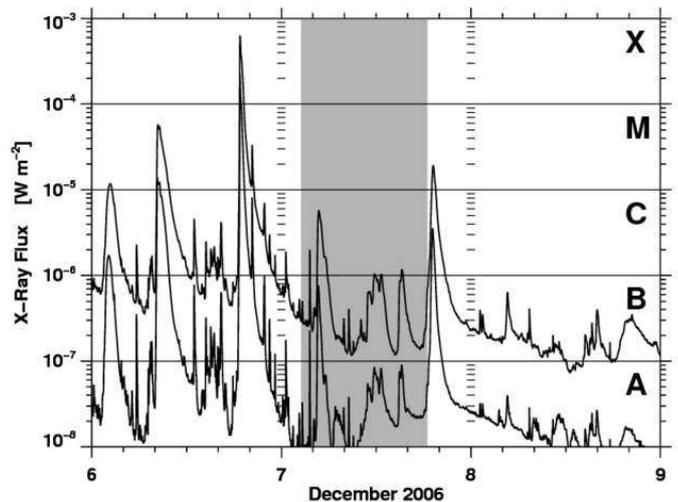}
\caption{One-minute GOES-11 X-ray flux on 2006 December~6--9 obtained in the
    0.05--0.40~nm (\textit{bottom}\tsp) and 0.1--0.8~nm (\textit{top}\tsp)
    energy channels. The shaded region indicates the time interval of 16~hours
    covered by \textit{Hinode} data analyzed in the present study.}
\label{FIG02}
\end{figure}

The region-of-interest (ROI) with a size of 40~Mm $\times$ 40~Mm was centered on
a small decaying sunspot in the western part of the active region (see the white
rectangular region in Fig.~\ref{FIG01} before geometrical correction). This
satellite spot significantly evolved over the course of 16~hours. Accompanied by
C- and M-class flares (Tab.~\ref{TAB01}), the penumbra of the small spot almost
completely vanished. The X-ray flux over three days as measured by the
Geostationary Operational Environmental Satellite (GOES) is shown in
Fig.~\ref{FIG02}, where the shaded region indicates the 16-hour observing
period. Before measuring horizontal proper motions, the signature of the
five-minute oscillation was removed from the images by using a three-dimensional
Fourier filter with a cut-off velocity of 8~km~s$^{-1}$, which corresponds
roughly to the photospheric sound speed. To scrutinize flow fields associated
with slow penumbral decay and its relationship to flaring, we used the LCT
method described in \citet{Verma2011}.The technique computes cross-correlations
over
$32 \times 32$-pixel regions with a Gaussian sampling window having a FWHM of 15
pixels (1200~km) corresponding to the typical size of a granule.

%###############################################################################
%#
%#    RESULTS
%#
%###############################################################################

\section{Results}

In the following, we will describe the decay of the satellite sunspot, compute
the photometric decay rates of its umbra and penumbra, and study the impact of
the decay process on photospheric and chromospheric horizontal flows. The
photometric observations together with a temporal analysis of the flow maps
yields estimates of decay times of various solar features. Finally, we relate
these findings to an M2.0 flare, which occurred towards the end of the 
time-series.

% FIGURE 2

% FIGURE 4
\begin{figure*}
\includegraphics[width=\textwidth]{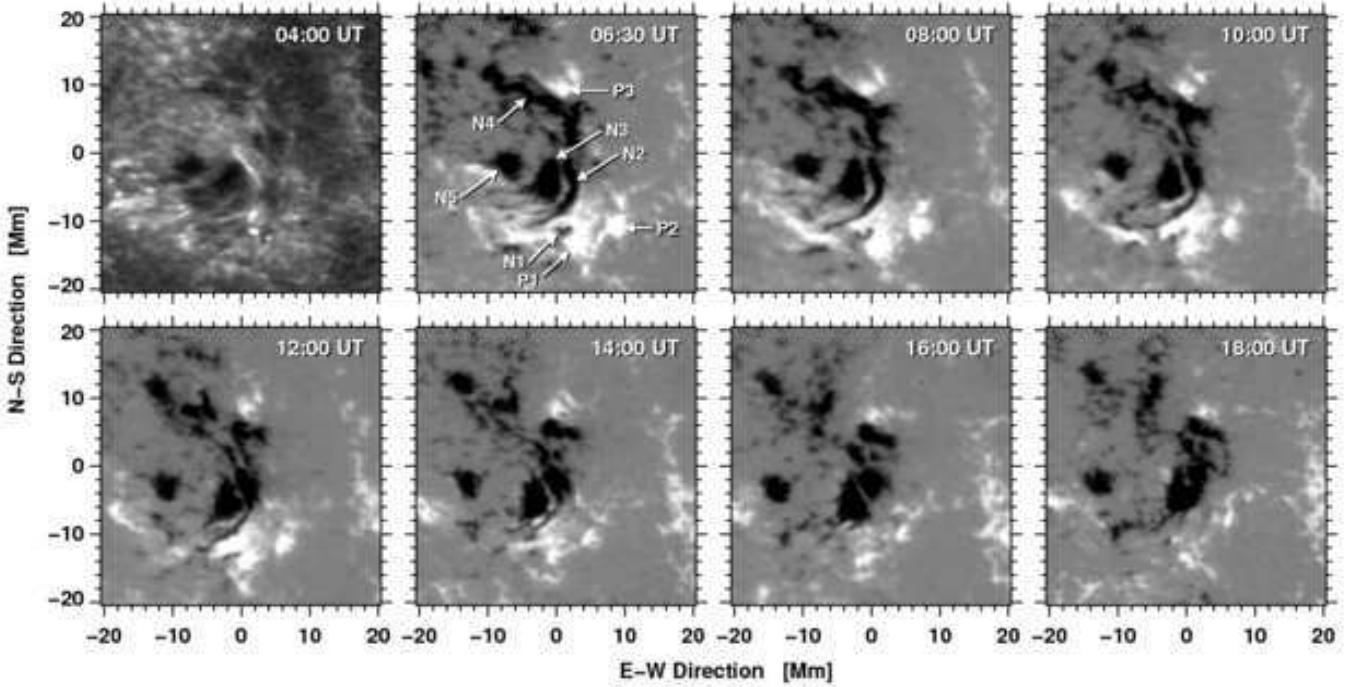}
\caption{Time-series of calibrated and geometrically corrected
    \textit{Hinode}/NFI magnetograms corresponding to the G-band images in
    Fig.~\ref{FIG03}. The line-of-sight
   magnetic field $B_{LOS}$ is displayed in the range of $\pm 800$~G as in
   Fig.~\ref{FIG01}.} 
\label{FIG04}
\end{figure*}

\subsection{Morphology\label{SEC03.1}}

The temporal evolution of the decaying satellite sunspot is shown at two-hour
intervals in Fig.~\ref{FIG03}. Note that after geometric correction the axis
labels no longer refer to heliographic coordinates on the solar disk. They are
more readily provided for easier identification and comparison of intensity or
flow features. As a convenient reference to photometric and magnetic components
of the satellite spot, we provided numbered labels \textsf{P} and \textsf{N} in
Figs.~\ref{FIG03} and \ref{FIG04} referring to positive- and negative-polarity
features, respectively. Initially, the ROI contains a few umbral cores with
rudimentary penumbrae, which are embedded in regions covered by G-band bright
points. The strongest umbral core \textsf{N3} is located close to the center of
the ROI, where it is separated from an elongated umbral core \textsf{N2} by a
faint light-bridge. The southern side of the light-bridge flares out in a strand
of penumbral filaments, which wind around the strong umbral core. Non-radial
penumbral filaments are indicative of shear flows \citep[e.g., ][]{Denker2007c,
Jiang2012}. These filaments became significantly weaker in response to a C6.1
flare at 04:45~UT, which can be considered as a form of a `rapid' penumbral
decay \citep{Wang2004, Liu2005}, however, in this case observed very localized
and at high spatial resolution. In addition to shear flows, we observed an
umbral core \textsf{N5} drifting away eastward from the dominant umbral core
\textsf{N3}. The separation of these umbral cores grew by 6~Mm over the course
of 14 hours, i.e., the separation speed is about 120~m~s$^{-1}$. Three small,
1-Mm wide pores of positive and negative polarity (\textsf{P1}, \textsf{P2}, and
\textsf{N1}) are located to the south of the strong umbral core. All three pores
disappeared by the end of the time-series. In addition, penumbral regions
started decaying across the entire ROI at about 10:00~UT. At 12:00~UT only weak
penumbral signatures were present so that only scattered pores were left by
14:00~UT. This period of `slow' penumbral decay was accompanied by several
B-class flares and a C1.1 flare at 11:48~UT.

In addition to the photometric evolution shown in Fig.~\ref{FIG03}, we also
trace in Fig.~\ref{FIG04} the reduction of magnetic complexity in the satellite
sunspot, which is based on \textit{Hinode} NFI magnetograms taken in the
photospheric Fe\,\textsc{i} $\lambda$630.25~nm line. The Stokes-$V/I$
magnetograms have $1024 \times 1024$~pixels and an image scale of
0.16\arcsec~pixel$^{-1}$. From the Stokes-$V$ and $I$ signal the approximate
line-of-sight magnetic field was calculated as 
\begin{equation} 
B_\mathrm{LOS} = - \frac{C_V}{0.798 C_I} \times 10^4~\mathrm{Mx~cm}^{-2} 
\end{equation} 
given in \citet{Isobe2007}. We scaled the magnetograms displayed in
Figs.~\ref{FIG01} and \ref{FIG04} using this relation.
However, no attempt was made to translate the LOS magnetic fields into the field
component $B_z$ normal to the solar surface. These data were only corrected for
geometrical foreshortening and carefully matched to the G-band images.
Unfortunately, no magnetograms were available until 06:30~UT so that we
substituted a Ca\,\textsc{ii}\,H image in the first panel of Fig.~\ref{FIG04},
and we had to resort to a magnetogram taken about 30~min later than the G-band
companion in the second panel. Note that the proximity of the active region to
the solar limb casts doubt on simply associating the sign of the circular
polarization with the sign of the magnetic fields component that is normal to
the solar surface. In particular, we observe in the NFI magnetograms at the
begining of the time-series apparent polarity reversals in limbward penumbral
regions for both the major and satellite sunspots, which are clearly related to
the close to horizontal magnetic field lines in these regions.

At the beginning of the time-series, the satellite sunspot was much more complex
than either the main spot or the developing $\delta$-spot. Four magnetic
features with negative polarity and three features with positive polarity play a
role in the decay of the satellite sunspot. Over a period of 16
hours the complexity of the magnetic fields was much reduced. The dominant umbra
\textsf{N3} is
separated by a light-bridge from a curved and elongated umbra \textsf{N2}.
Brightening of the small-scale features along the light-bridge around 06:00~UT
can be taken as an indication that convection penetrates the strong magnetic
fields, thus, contributing to the decay of the satellite spot. During this
decay the satellite sunspot slowly rotates counterclockwise ($\approx
3.6\arcdeg$~h$^{-1}$) as can be seen in Fig.~\ref{FIG02} using the light-bride
as a tracer. This slow rotation adds up to more than 50\arcdeg\ over the course
of the observations.

The core of the satellite spot, i.e., \textsf{N2}, \textsf{N3}, and the
light-bridge, survive until the end of the time-series. The curved, non-radial
penumbral filaments associated with this core can be traced not just in
intensity but also in magnetograms. The observed alternating polarities of the
filaments could just be projection effects of nearly horizontal magnetic fields.
This effect is likewise visible at the limb-side penumbra of the main spot and
for the magnetic element \textsf{N5}. Despite of its strong proper motion (see
above), the photometric and magnetic decay of \textsf{N5} is slow. Only towards
the end of the time-series the pore starts to break up. A conspicuous, bright
umbral dot appears in the center of the pore around 18:00~UT. Together with two
protrusions from the periphery of the pore, it forms a structure like a
light-bridge indicating further erosion of the pore. This erosive process is
easier to follow in intensity than in the magnetic field evolution, where
\textsf{N5} remains compact.

Two small pores \textsf{N1} and \textsf{P1} are located to the south of the
dominant umbra. Magnetic cancellation characterizes the evolution of this small
magnetic bipole. As the minority polarity \textsf{N1} disappeared earlier than 
\textsf{P1}. At 10:00~UT only \textsf{P1} was left. In general, the photometric
decay proceeds faster than the decline of the magnetic flux. This also holds
true for the third small pore \textsf{P2}, which survived somewhat longer than
\textsf{P1} and can still be detected as a tiny magnetic knot in the G-band
image at 18:00~UT.

To the east of a small umbral core \textsf{P3} we observe a rudimentary penumbra
of opposite polarity, which had dissolved by 14:00~UT. The underlying magnetic
field \textsf{N4} of the rudimentary penumbra was elongated stretching over a
distance of about 18~Mm connecting to the also elongated umbral core
\textsf{N2}. Together, \textsf{N2} and \textsf{N4} define the magnetic neutral
line of the satellite sunspot, with strong positive fields at the extremities to
the south (\textsf{P1} and \textsf{P2}) and  north (\textsf{P3}). The slow
cancellation of flux associated with \textsf{P3} continues over the 16-hour
period and results in long-lived flow features in the decorrelation maps (see
Sect.~\ref{SEC03.4}).

% FIGURE 5
\begin{figure}
\includegraphics[width=\columnwidth]{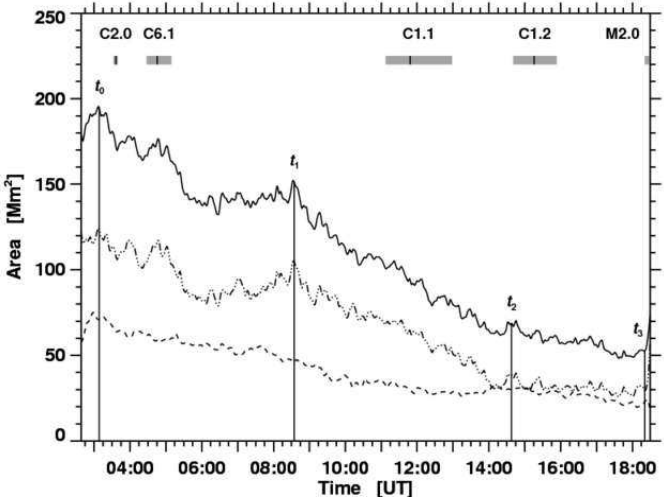}
\caption{Temporal evolution of the area covered by active region NOAA~10930 over
    the 16-hour period from 02:30--18:30~UT. The area enclosed by umbral cores
    and pores is displayed with a dashed line. The dash-dot-dotted line refers
    to the penumbra, whereas the solid line denotes the total area. The times
    $t_0$, $t_1$, $t_2$, and $t_3$ delimit time intervals over which the linear
    photometric decay rates were computed (see Tab.~\ref{TAB02}). The gray bars
    depict the duration of flares (Tab.~\ref{TAB01}), where the thin black line 
    marks the peak time for flares.}
\label{FIG05}
\end{figure}

% TABLE 2
\begin{table}
\caption{Decay rates [Mm$^{2}$~day$^{-1}$] of penumbra, umbra, and the
entire satellite sunspot (pores, umbrae, and penumbrae) for four time
intervals.}\label{TAB02}
\tiny
\begin{center}
\begin{tabular}{lcccc}
\hline\hline
Feature & \multicolumn{4}{c}{Decay rate [Mm$^{2}$~day$^{-1}$] for four time
intervals}\rule[-2mm]{0mm}{5mm}\\
        & $t_{0}-t_{1}$ & $t_{1}-t_{2}$ & $t_{2}-t_{3}$ &
$t_{0}-t_{3}$\rule[-2mm]{0mm}{3mm}\\
\hline
Penumbra        & $130.2$ & $244.2$ & $31.3$ &  \rule[-1mm]{0mm}{4mm}\\
Umbra           & \phn$86.4$ & \phn$68.4$ & $67.9$ &
\phn$73.7$\rule[-1mm]{0mm}{4mm}\\
Spot            & $216.6$ & $312.6$ & $99.3$ & $225.8$\rule[-1mm]{0mm}{4mm}\\
\hline
\end{tabular}\\
\tiny \hspace*{0.3mm}
\parbox{65mm}{\vspace*{-2mm} 
\begin{itemize}
\item[Note:] The time intervals are $t_{0}$= 03:07~UT, $t_{1}$= 08:36~UT,\\
$t_{2}$= 14:46~UT, $t_{3}$= 18:20~UT.
\end{itemize}}
\end{center}
\end{table}

\subsection{Decay rates\label{SEC03.2}}

We used morphological thresholding for discerning various solar features in
intensity. Images were normalized using eight quiet Sun regions evenly spread
over the time-series. Thus, a weak trend in mean intensity (center-to-limb
variation related to solar rotation) was removed from the data. We used a fixed
intensity threshold of $I_\mathrm{mag} < 0.85$ for strong magnetic features and
an adaptive threshold for G-band bright points \citep[see][]{Verma2011}, which
is given as 
\begin{equation}
I_\mathrm{bp} = 1.15 + 0.2(1-\mu)\,.
\end{equation}
The umbrae were identified using a fixed threshold of $I_\mathrm{dark}=0.6$,
while sunspot penumbrae cover intermediate intensities from $I_\mathrm{mag}$ to
$I_\mathrm{dark}$, and granulation lies between $I_\mathrm{mag}$ to
$I_\mathrm{bp}$. The adaptive threshold was used as a first-order approximation
to take into account the center-to-limb variation of G-band bright points
because they exhibit an enhanced contrast near the solar limb. The thresholding
algorithm also allows us to compute the area for different solar features. The
algorithm provides in general a good estimate of the area except for a few small
features and towards the end of the sequence, when only pores are left, for
which the borders are erroneously classified as penumbra. If many small pores
are present, then their periphery with intensities like penumbra can be
significantly larger than the cores of the pores themselves.

The curves in Fig.~\ref{FIG05} represent the temporal evolution of areas
subsumed by umbral cores/pores, penumbrae, and the entire satellite spot (sum of
pores, umbra, and penumbra). At the beginning of the time-series, the area of
all magnetic features was about 200~Mm$^2$, of which approximately two thirds
were classified as penumbra and the remaining third corresponded to umbral
cores/pores. By the end of the time-series only the later features stayed on
covering little over 50~Mm$^2$, i.e., only one quarter of the area intially
covered by the strong magnetic features.  The features' decay rates were
computed using linear regression. We divided the time-series into four time
intervals $t_{0}-t_{1}, t_{1}-t_{2}, t_{2}-t_{3}, $ and $t_{0}-t_{3}$, because
the penumbral decay curve has two peaks at t$_0$ and t$_1$, and the penumbra had
completely vanished at $t_2$. The time $t_3$ was chosen just eight images before
the end of the time-series so that artifacts from the subsonic filtering become
negligible, and it marks the onset of the M2.0 flare.

Slow penumbral decay is the distinctive
feature of the satellite sunspots's temporal evolution. Table~\ref{TAB02}
provides the respective photometric decay rates for the above mentioned time
intervals. The penumbral decay rate significantly increases from $t_0$ -- $t_1$
to $t_1$ -- $t_2$ almost doubling its value to 244.2~Mm$^2$~day$^{-1}$. A linear
regression might not be the best choice for the penumbral decay rate (130.2
Mm$^2$ day$^{-1}$) during the first time interval because there is a sudden
decay in the area after the C6.1 flare at 04:45~UT. During the time interval 
$t_2$ -- $t_3$ the penumbral area decayed by  4.7~Mm$^2$ in 3.6~h, which
corresponds to a decay rate of 31.3~Mm$^2$~day$^{-1}$. However, this is just an
artifact of the intensity thresholding algorithm used in classifying penumbral
areas, which erroneously labels the boundaries of pores as penumbra. The
1-$\sigma$ uncertainties for the decay rates are about 2.5~Mm$^2$~day$^{-1}$.

Therefore, computing a
penumbral decay rate for the entire time period
$t_0$ -- $t_3$ is not meaningful. The umbral decay rates do not vary much. The
overall value of 73.7~Mm$^{2}$~day$^{-1}$ during $t_0$ -- $t_3$ represents
closely the decay rates for the shorter intervals. Hence, the penumbra decayed
three time faster as the umbral cores. The decay rate of the spot is just the
sum of the umbral and penumbral decay rates. The overall spot decay rate in
current study (225.8~Mm$^{2}$~day$^{-1}$ for $t_0$--$t_3$) is well within the
range of previously reported values, e.g., \citet{Bumba1963} found a decay rate
of 180~Mm$^{2}$~day$^{-1}$ for non-recurrent groups. Spot decay rates in other
studies range from 10--125~Mm$^{2}$~day$^{-1}$
\citep[e.g.,][]{Moreno-Insertis1988, Pillet1993, Hathaway2008}.

\subsection{Flow fields in photosphere and chromosphere\label{SEC03.3}}

% FIGURE 6
\begin{figure*}
\includegraphics[width=\textwidth]{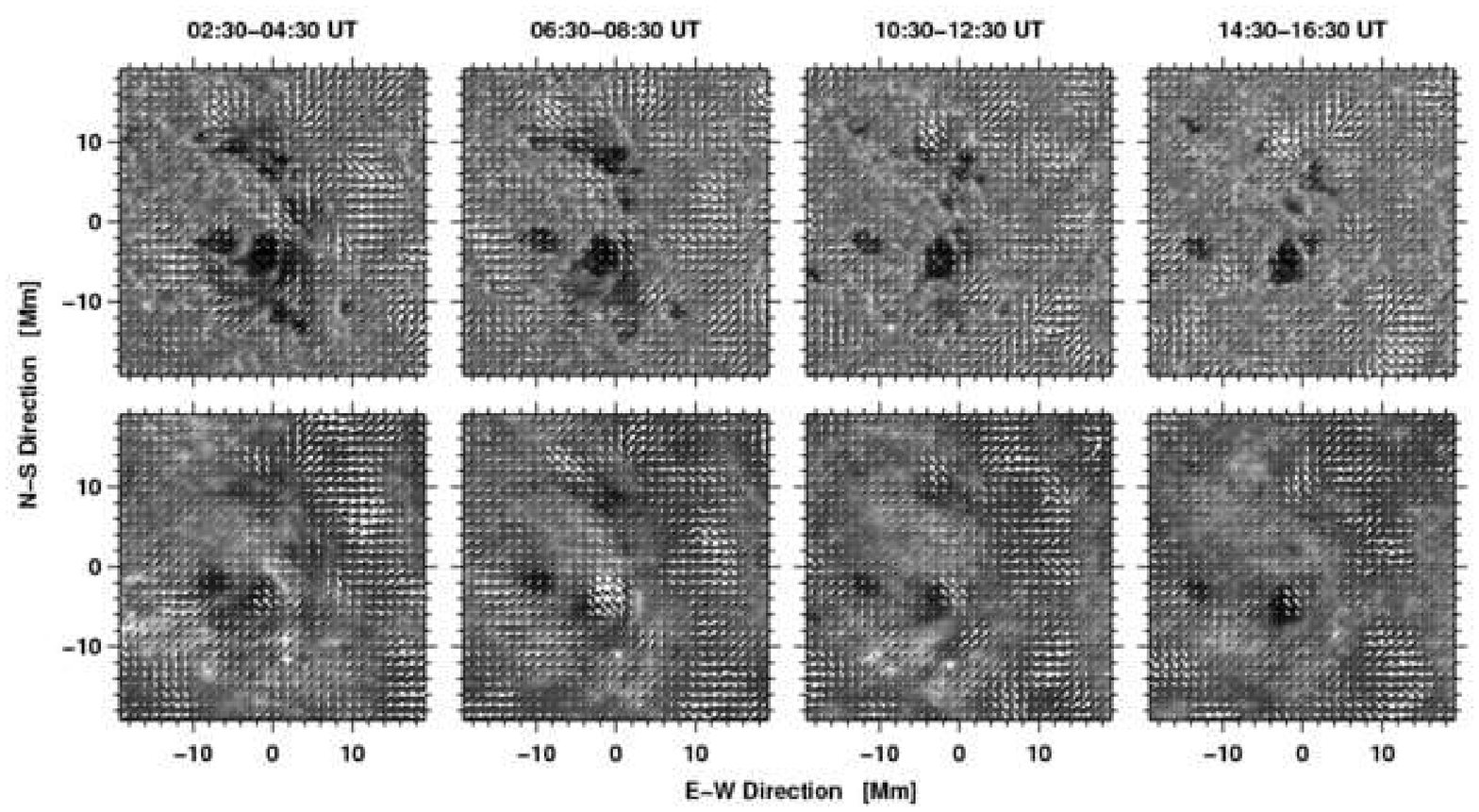}
\caption{Horizontal flow fields averaged over two hours are depicted as
    arrows overlaid on geometrically corrected G-band (\textit{top}\tsp) and
    Ca\,{\sc ii}\,H (\textit{bottom}\tsp) images of four time-series. The
    arrows indicate magnitude and direction of the horizontal proper motions.
    Arrows with a length corresponding to the grid size indicate velocities of
    0.5~km~s$^{-1}$.}
\label{FIG06}
\end{figure*}

% TABLE 3

\begin{table*}
%\parbox{1.25\columnwidth}
\caption{Parameters describing horizontal proper motions for various solar
     features based on the four G-band and Ca\,{\sc ii}\,H time-series with
     $\Delta T=2$~h and $\Delta t=60$~s for four time-series shown in
     Fig.~\ref{FIG06}.}\label{TAB03}
\tiny
\begin{tabular}{llccccc}
\hline\hline
     & &
    $\bar{v}$ &
    $v_\mathrm{med}$ &
    $v_{10}$ &
    $v_\mathrm{max}$ &
    $\sigma_{v}$\rule[-2mm]{0mm}{5mm}\\
Feature       & Image Type      & [km~s$^{-1}$] & [km~s$^{-1}$] & [km~s$^{-1}$] & [km~s$^{-1}$] &  [km~s$^{-1}$]\rule[-2mm]{0mm}{3mm}\\
\hline
All           & G-Band          & $0.33$ & $0.29$ & $0.63$ & $1.46$ &  $0.21$\rule{0mm}{4mm}\\
              & Ca\,{\sc ii}\,H & $0.35$ & $0.28$ & $0.69$ & $2.31$ &  $0.25$\\
Granulation   & G-Band          & $0.35$ & $0.31$ & $0.65$ & $1.43$ &  $0.21$\rule{0mm}{4mm}\\
              & Ca\,{\sc ii}\,H & $0.36$ & $0.30$ & $0.71$ & $2.31$ &  $0.25$\\
Penumbra      & G-Band          & $0.28$ & $0.24$ & $0.51$ & $1.22$ &  $0.19$\rule{0mm}{4mm}\\
              & Ca\,{\sc ii}\,H & $0.30$ & $0.24$ & $0.60$ & $1.36$ &  $0.21$\\
Umbra         & G-Band          & $0.20$ & $0.19$ & $0.33$ & $0.60$ &  $0.10$\rule{0mm}{4mm}\\
              & Ca\,{\sc ii}\,H & $0.30$ & $0.23$ & $0.66$ & $1.19$ &  $0.23$\\
Bright Points & G-Band          & $0.18$ & $0.17$ & $0.32$ & $0.71$ &  $0.10$\rule{0mm}{4mm}\\
              & Ca\,{\sc ii}\,H & $0.19$ & $0.18$ & $0.32$ & $0.82$ &  $0.10$\rule[-2mm]{0mm}{0mm}\\
\hline
\end{tabular}
\\
\tiny \hspace*{112mm}
\raisebox{10.5mm}[-10.5mm]{\parbox{71mm}{
%\begin{itemize}
%\item 
For computing the characteristic parameters of the penumbral flow field
we neglected the last time interval $t_2$--$t_3$, since by that time
the penumbra had decayed and the features detected as penumbra were just
artifacts of the thresholding algorithm. The time period $t_0$--$t_2$ was
excluded from calculating Ca\,\textsc{ii}\,H flow parameters of the umbra, since
motions along post-flare loops resulted in high flow speeds. 
%\end{itemize}
}}
\end{table*}

The two-hour averaged LCT flow maps shown in Fig.~\ref{FIG06} were computed
using the 16-hours time-series of \textit{Hinode} G-band and Ca\,\textsc{ii}\,H
images. We used the satellite sunspot as a reference and aligned all images in
time-series accordingly. These flow maps provide insight into horizontal proper
motions in the photosphere and chromosphere. We quantified horizontal proper
motions for various solar features (e.g., bright points, granulation, sunspot
penumbrae, and strong magnetic features) by applying morphological and adaptive
thresholds to G-band images (see Sect.~\ref{SEC03.1}). We applied the same
indexing to the Ca\,\textsc{ii}\,H flow maps to have an one-to-one
correspondence comparing photospheric and chromospheric flow fields, while
neglecting morphological differences. We calculated the mean $\bar{v}$, median
$v_\mathrm{med}$, maximum $v_\mathrm{max}$, $10^\mathrm{th}$ percentile
$v_{10}$, and standard deviation $\sigma_v$ of the horizontal flow speeds (see
Tab.~\ref{TAB03}). The standard deviation in the flow speed refers to the
variance in the data rather than to a formal error. Typical flow characteristics
of solar features were reported by \citet{Verma2011}, who also presented values
for other time intervals $\Delta T$ and cadences $\Delta t$. Here, we computed
the flow parameters for four time-series with $\Delta T=2$~h and afterwards
averaged them. The average flow parameters of the various solar features are
within the expected ranges except for granulation where $\bar{v} =
0.35 \pm 0.21$~km~s$^{-1}$, which is lower than the value of
$\bar{v} = 0.47 \pm 0.27$~km~s$^{-1}$ mentioned in
\citet{Verma2011}. However, the granular flow speed could be lower because of
the presence of strong magnetic flux concentrations. Regions of granulation,
which are not in close proximity to strong magnetic fields, often showed strong
divergence centers, e.g., in the south-west corner of the FOV towards the end of
the time-series. In general, photospheric and chromospheric flow parameters are
virtually the same. The only notable dissimilarity between the G-band and
Ca\,\textsc{ii}\,H flow maps relates to post-flare loops straddling the dominant
pore \textsf{N3}. These loops are most prominent in Ca\,\textsc{ii}\,H images
and result in strong flows of more than 1.14~km~s$^{-1}$ (Fig.~\ref{FIG06}:
06:30--08:30~UT). The signature of these post-flare loops is still visible at
later time periods. However, the associated flows are much weaker. Other than
that differences in the frequency distributions exist only for the high-speed
tail, i.e., the maximum speed $v_\mathrm{max}$ and to a lesser extend the
10$^\mathrm{th}$ percentile speed $v_{10}$.

In the two-hour averaged LCT flow maps the overall impression of flow vectors
for G-band and Ca\,\textsc{ii}\,H images is indistinguishable. The flow patterns
around the satellite sunspot are different from regular sunspots
\citep[e.g.,][]{Balthasar2010} because of its non-radial penumbra and location
within a complex active region. The most intriguing feature in the G-band flow
maps is the anticlockwise spiral motion around the dominant umbral core
\textsf{N3}. The light-bridge between \textsf{N2} and \textsf{N3} marks the
location of shear flows, where stronger flows ($\approx 0.45$~km~s$^{-1}$)
linked to the elongated umbra \textsf{N2} move past weaker flows($\approx
0.17$~km~s$^{-1}$) spiraling around the dominant umbra \textsf{N3}. The spiral
motion and shear flows were most conspicuous in the first flow map
(02:30--04:30~UT) but faded out once the penumbra had decayed. Strong outward
motions were present at the outer tips of penumbral filaments associated with
\textsf{N3} and \textsf{P3} in the northern part of the FOV. These strong
outward motions associated with \textsf{P3} continue to exist until the end of
the time-series and appear as long-lived features in the decorrelation maps of
the flow speed (see Sect.~\ref{SEC03.4}).

\subsection{Decorrelation times\label{SEC03.4}}

The lifetime of solar features can be estimated by selecting a reference map at
an instance $t_i$ and by computing how consecutive maps decorrelate with time.
As put forward by \citet{Welsch2012}, we computed linear Pearson and rank-order
Spearmann correlation coefficients for G-band intensity and the corresponding
horizontal flow speed. The flow speed maps were computed as sliding one-hour
averages centered on each point in time $t_i$. Pearson's correlation coefficient
$\rho$ measures the strength of a linear dependence of two variables $x$ and
$y$, and is computed by dividing the covariance of the two variables by the
product of their standard deviations:

\begin{equation}
\rho = \frac{\displaystyle
    \sum_{i=1}^N \left(x_i - \bar{x}\,\right)
                 \left(y_i - \bar{y}\,\right)}{\displaystyle
    \sqrt{\sum_{i=1}^N \left(x_i - \bar{x}\,\right)^2 \cdot
          \sum_{i=1}^N \left(y_i - \bar{y}\,\right)^2}}
\end{equation}
\begin{displaymath}
\mathrm{with}\;\,
\bar{x} = \frac{1}{N} \sum_{i=1}^N  x_i\;\,\mathrm{and}\;\,
\bar{y} = \frac{1}{N} \sum_{i=1}^N  y_i\,.
\end{displaymath}
The Spearman correlation coefficient $r_s$ is similarly defined as the Pearson
correlation coefficient between two ranked variables $\mathrm{r}(x)$ and
$\mathrm{r}(y)$. We used an IDL algorithm based on the recipe  provided in
\citet{Press1992}:
\begin{equation}
r_s = \frac{\displaystyle
    \sum_{i=1}^N \left(\mathrm{r}(x_i) - \overline{\mathrm{r}(x)}\,\right)
                 \left(\mathrm{r}(y_i) - \overline{\mathrm{r}(y)}\,\right)}{
                 \displaystyle
    \sqrt{\sum_{i=1}^N \left(\mathrm{r}(x_i) -
                             \overline{\mathrm{r}(x)}\,\right)^2 \cdot
          \sum_{i=1}^N \left(\mathrm{r}(y_i) -
                             \overline{\mathrm{r}(y)}\,\right)^2}}
\end{equation}
\begin{displaymath}
\mathrm{with}\;\,
\bar{x} = \frac{1}{N} \sum_{i=1}^N \mathrm{r}(x_i)\;\,\mathrm{and}\;\,
\bar{y} = \frac{1}{N} \sum_{i=1}^N \mathrm{r}(y_i)\,.
\end{displaymath}

The 16-hour time-series was divided into two parts. The first eight-hour
time-series covers the phase of penumbral decay, whereas towards the second half
the penumbra had already vanished. Autocorrelation functions $r_s (t)$ were
computed for intensity and flow maps over circular areas with a diameter of
about 5~Mm and for lag times of up to 300~min. We used 16 and 10 reference
frames for the intensity and speed maps, respectively. The number of reference
frames had to be reduced to 10 in case of the flow maps because of the 60-minute
sliding average. We averaged the autocorrelation functions to have different
instances of surface features contributing to our sample. In both cases, the
averages are based on 460~min of data, i.e., either $16 \times 10 + 300$~min or
$10 \times 10 + 300 + 60$~min, where the last term results from the sliding
averages. The thus averaged autocorrelation functions are fitted with decaying
exponential functions.

\begin{eqnarray}
r_s (t) & = & \Big[     r_s \big(f(t_0), f(t_1)\big),\;
                        r_s \big(f(t_0), f(t_2)\big),\; \ldots\nonumber\\
        &   & \ldots,\; r_s \big(f(t_0), f(t_{n-1})\big),\;
                        r_s \big(f(t_0), f(t_n)\big) \Big] \\
        & \approx & \exp\left(-\frac{1}{\tau} t^\gamma\right)\nonumber \,,
\end{eqnarray}
where the constants $\tau$ and $\gamma$ are derived from a least-squares fit
to the measured $r_s (t)$. The function $f(t)$ is either an intensity or
a horizontal flow speed map. The lifetimes (or decorrelation times) were
calculated for $r_s(t) = 1/2$, i.e.,
\begin{equation}
t_{\nicefrac{1}{2}} = (\tau\ln 2)^{\nicefrac{1}{\gamma}}\,.
\label{EQN06}
\end{equation}

% FIGURE 7
\begin{figure*}
\includegraphics[width=\textwidth]{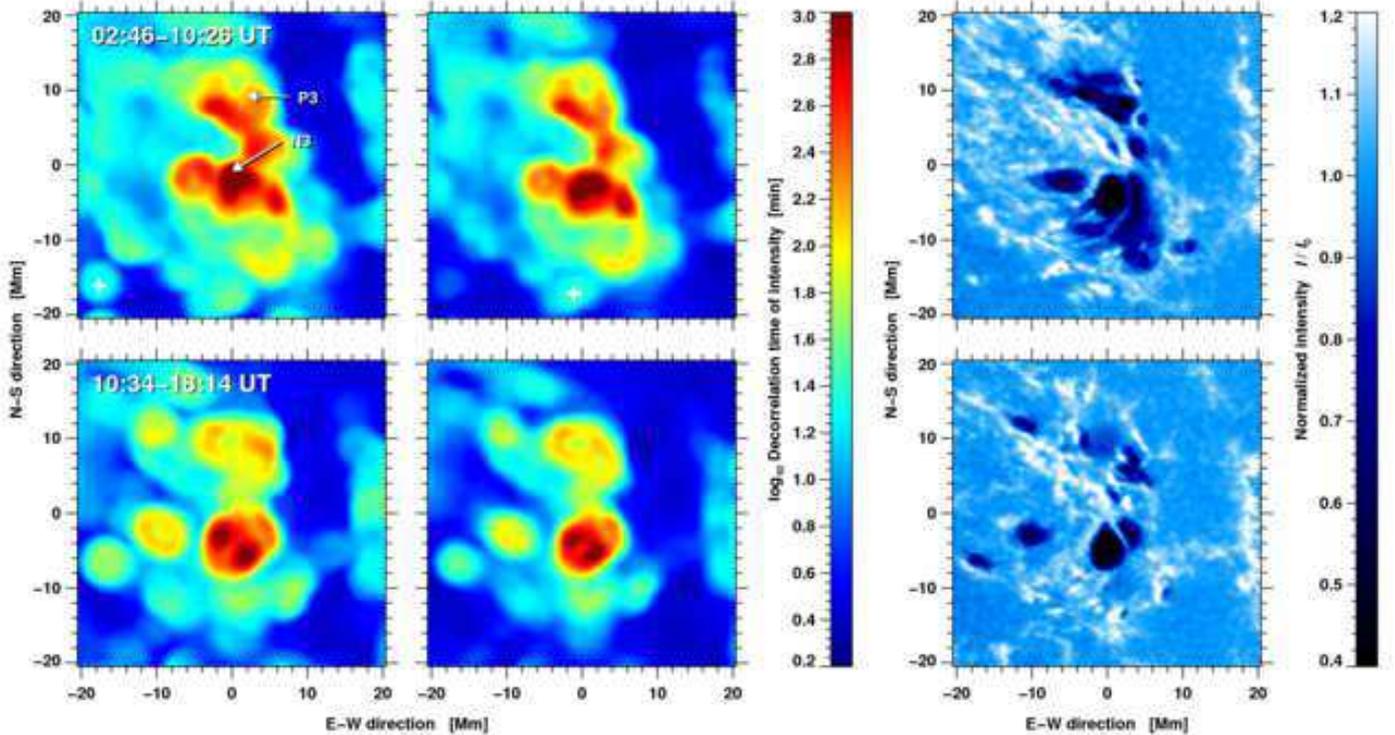}
\caption{Decorrelation times computed using the linear Pearson (\textit{left})
    and Spearman's rank order (\textit{middle}) correlation coefficients for the
    intensity of G-band images. The time-series was
    divided into two parts from 02:46--10:26~UT (\textit{top}) and
    10:34--18:14~UT (\textit{bottom}). Average G-band images for these time
    intervals (\textit{right}) clearly show the penumbral decay of the satellite
    sunspot.}
\label{FIG07}
\end{figure*}

The case $\gamma = 1$ corresponds to a simple exponential decay and $\tau$
would simply refers to the lifetime of the feature. This is the standard
approach determining lifetimes of solar granulation but this simple decay law
no longer holds in the presence of (strong) magnetic fields. Using also $\gamma$
as a free fit parameter results in a much improved
$\chi^{2}$-statistics independent of the magnetic environment. If $\gamma \ne
1$, then $\tau$ can no longer be interpreted as lifetime. Therefore, we chose
Eqn.~\ref{EQN06} to determine how long solar features last because it
removes the entanglement of $\tau$ and $\gamma$ by emphasizing the quality of
the fit. 

A more detailed analysis shows that $\gamma \approx 1$ in
regions with granulation and $\gamma \approx 1/2$ in sunspots are
representative values for the autocorrelation functions of intensity features.
The autocorrelation functions of flow features are characterized by 
$\gamma \approx 3/2$ with no major differences between sunspots and granulation.
However, the exponent can approach $\gamma \approx 3$ for some long-lived
features, e.g., stationary G-band bright points. In these cases, the $\tau$
values become very large and can no longer be interpreted as a lifetime.

\begin{figure*}
\includegraphics[width=\textwidth]{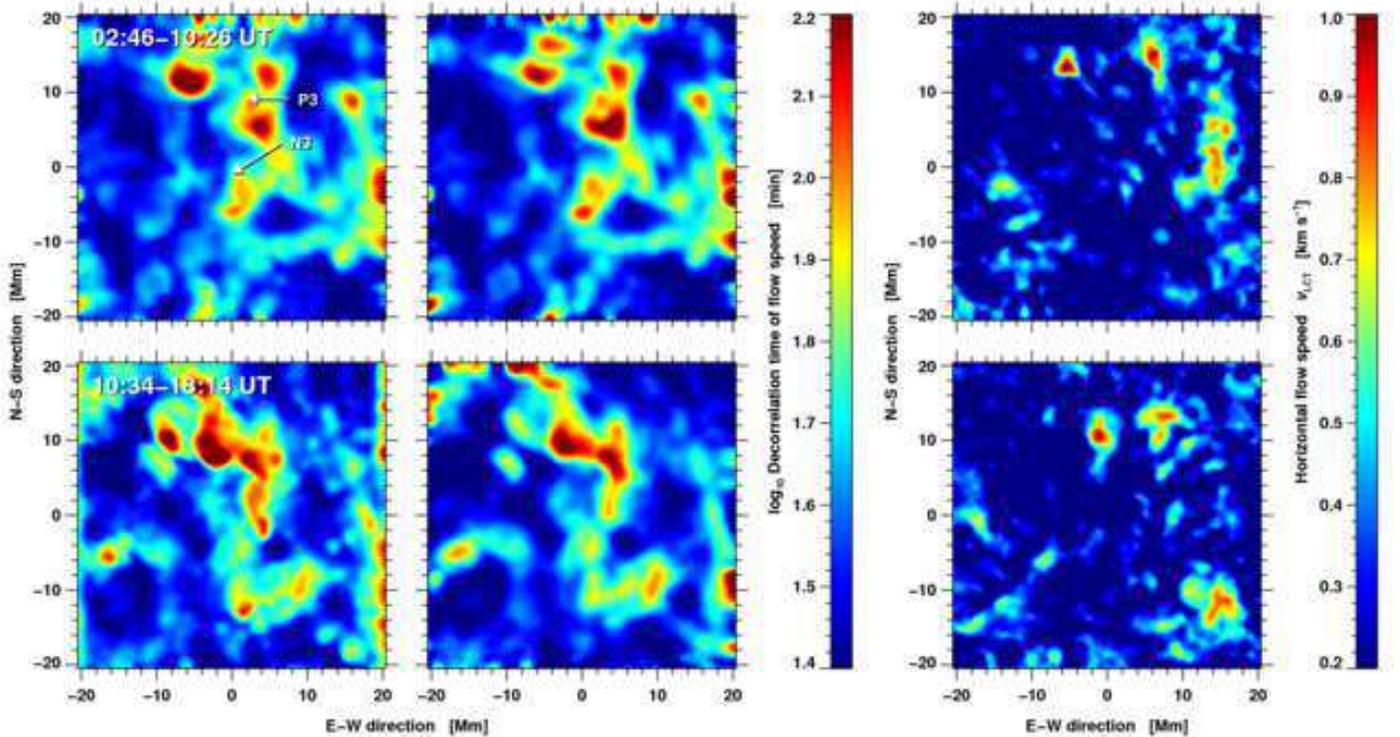}
\caption{Decorrelation times computed using the linear Pearson (\textit{left})
    and Spearman's rank order (\textit{middle}) correlation coefficients for the
    horizontal flow speed. The time-series was divided into two parts from
    02:46--10:26~UT (\textit{top}) and 10:34--18:14~UT (\textit{bottom}).
    Average
    flow speed maps (\textit{right}) show several small regions with enhanced
    flows in the vicinity of long-lasting flow features.}
\label{FIG08}
\end{figure*}

In Figs.~\ref{FIG07} and \ref{FIG08}, we depict the decorrelation-time maps for
the intensity and flow speed, respectively. Linear and rank-order correlations
qualitatively produce the same decorrelation times. Our type of presentation
differs from \citet{Welsch2012} because our decorrelation times were computed
either for each pixel in the FOV (Pearson's $\rho$) or for a $4\times4$-pixel
grid (Spearman's $r_s$). The coarser sampling is necessitated because of the
computational overhead in ranking the variables. These maps were then smoothed
by a Gaussian with a FWHM of 1200~km. We inferred from the decorrelation maps of
the intensity that the magnetic features have longer lifetimes than granulation,
which is expected.

Regions of granulation can be extracted from the average G-band images shown in
the right panels of Fig.~\ref{FIG07}, where they correspond to the dark blue
regions to the west of the satellite sunspot. Typical granular lifetimes derived
from the linear and rank-order decorrelation maps are about 3.1~min and 4.6~min,
respectively. In the rank-order case, we find a standard deviation of the
frequency distributions of about 1~min. Because of a pronounced tail towards
longer lifetimes, the standard deviation of the linear case is much higher.
However, the median is close to 3~min in both cases. There is no significant
difference in the granular lifetimes during the first and second halves of the
time-series. Our findings are in good agreement with the original work of
\citet{Bahng1961} and match accurately the findings of \citet{Title1989}.

G-band bright points show two types of behavior: either they remain stationary
as at the supergranular boundary at the western periphery of the FOV or they
stream away from the satellite sunspot towards the main spot. In the
long-duration G-band images the bright point coalesce into strands indicating
preferential paths taken by small-scale magnetic features. Similar motion
patterns were observed for the moat flow of decaying sunspot \citep{Verma2012}.
However, thread-like concentrations of magnetic field were also observed by
\citet{Strous1996} for an emerging active region. Thus, preferential paths for
the migration of small-scale magnetic elements might be a common property of
flux emergence, removal, and dispersal. The factor that G-band bright points
being either stationary or moving does not have an impact on their lifetimes.
The light-blue to turquoise colors denote lifetimes of about 35~min in linear
and 25~min in rank-order decorrelation maps. The distributions in both cases
have a pronounced high-lifetime tail. Note that individual, long-lived, and
stationary G-band bright points can create artifacts, which appear as
disk-shaped features (e.g., marked with white $+$ in Fig.~\ref{FIG07})
in the smoothed decorrelation maps because as long as they
are fully contained in the circular correlation window, their presence outweighs
any other contribution to the correlation function.

% FIGURE 8

The typical lifetime of strong magnetic features (pores, umbrae, and penumbrae)
are about 200~min and 235~min using linear and rank-order correlations. There is
virtually no difference in the respective distributions of magnetic features,
except that the area covered by extremely long-lived feature (lifetimes $>$
500~min) is moderately higher (1.8~Mm$^2$) in case of rank-order correlations as
compared to linear correlations (1.2~Mm$^{2}$). The photometric decay of the
satellite sunspot also leaves its mark in the decorrelation maps. The area
covered by features living longer than 100~min decreased from  160~Mm$^2$ to
80~Mm$^2$, while in parallel the complexity of the long-lived features
reduced. The lifetime in the vicinity of the umbral core \textsf{P3} was about
300~min during the first half of the time-series. In the second half erosion of
the rudimentary penumbra and slow cancellation of magnetic flux near \textsf{P3}
led to significantly shorter lifetimes (about 100~min) in that region. The only
conspicuous feature that remained in the latter decorrelation map is a compact
oval region associated with the dominant umbral core \textsf{N3}. In two small
kernels with an area of about 2~Mm$^{2}$ the lifetime exceeds 1000~min.

Decorrelation maps of the flow speed are presented in Fig.~\ref{FIG08}. The
right panels show the averaged, long-duration flow maps. Flow speeds are
suppressed in regions containing strong and weak magnetic fields. Typical values
are $0.24\pm0.15$~km~s$^{-1}$, where the standard deviation refers to the
variance within the region covered by the satellite sunspot and the surrounding
G-band bright points. Higher flow speeds ($0.52\pm0.11$~km~s$^{-1}$) are
encountered in quieter areas. In particular in proximity to supergranular
boundaries near the western periphery of the FOV, flow speeds approaching
1~km~s$^{-1}$ are measured. For example, a strong divergence region (see
Fig.~\ref{FIG06} at time periods 10:30--12:30~UT and 14:30--16:30~UT) becomes
apparent in the lower-right corner of the FOV in the second half of the
time-series.

Two small flow kernels of about 1~Mm$^2$ possessing high-speed values
approaching 1~km~s$^{-1}$ are related to the decay of the rudimentary penumbra
near the umbral core \textsf{P3}. They are related to slow flux cancellation in
that region. These flow kernels also leave an imprint in the decorrelation maps.
Note that the logarithmic scale of the decorrelation times significantly differs
for G-band intensities and horizontal flow speeds. Surprisingly, the lifetimes
associated with the dominant umbral core \textsf{N3} are much lower than the
ones for \textsf{P3}, i.e., horizontal flows are more persistent near
\textsf{P3}. Thus, flows contribute to the decay of the satellite sunspot most
noticeably in regions with weaker and less compact magnetic fields. Persistent
flows have lifetimes typically above 80~min and survive up to 160~min. The
frequency distributions of flow lifetimes derived from linear and rank-order
decorrelation maps are virtually same. In both cases, the distributions are
broad and do not have a high-velocity tail. Comparing Figs.~\ref{FIG07} and
\ref{FIG08}, a higher fine-structure contents becomes immediately apparent,
which might be explained by the shorter lifetimes of horizontal flow patterns.

\subsection{Homologous M2.0 flare\label{SEC03.5}}

An M2.0 flare started in active region NOAA~10930 at 18:20~UT towards the end of
the time-series. Time-resolved Ca\,\textsc{ii}\,H flare  kernels are shown in
the right panel of Fig.~\ref{FIG01}. The color-coded flare kernels are based on
images with the full cadence of 30~s. The first indications of this two-ribbon
flare were associated with the remnants of the negative polarity features around
\textsf{N2} and \textsf{N3} in the satellite sunspot. The second ribbon is
located some 10--15~Mm away in a network region to the west with positive
polarity. Flare brightenings related to the $\delta$-spot appear delayed by a
few minutes. Furthermore, remote brightenings appear in a negative-polarity
Ca\,\textsc{ii}\,H plage region towards the north-west. A small filament above
the neutral line formed by the remnants of \textsf{N2}/\textsf{N3} and
\textsf{P1}/\textsf{P2} erupted during the flare. The location of this filament
was extracted from BBSO H$\alpha$ full-disk images and carefully matched with
the magnetogram of Fig.~\ref{FIG01}.

The X6.5 flare on 2006 December~6 was also a two-ribbon flare. Here, the flare
ribbons were located inside the umbra of the main spot and along the neutral
line separating the main and $\delta$ spots. A filament located above this
neutral line \citep[see F1 in Fig.~11 of][]{Bala2010} erupted as a result of the
flare and produced an impressive Moreton wave. The latter flare ribbon extended
all the way to the satellite sunspot, which at that time was still connected to
the main spot by a wide band of penumbral filaments. Rapid penumbral decay
initiated by the X6.5 flare characterize this penumbral region. \citet{Deng2011}
find no indications for flux emergence in this regions but attribute the
initiation of the flare to (shear) flows along the magnetic neutral line, which
were enhanced just before the onset of the flare. The rapid penumbral decay
within the band connecting main and satellite sunspots \citep{Wang2012} also
marks the beginning of demise of the satellite sunspot.

Even though being separated by more than one order of magnitude in
X-ray flux, the X6.5 and M2.0 flare share a variety of traits, so that they can
be considered as homologous flares: both are two ribbon flares, filament
eruption are observed in both cases, the reconfiguration of the magnetic field
topology involves the satellite sunspots, and flux removal rather than
emergence is a decisive mean in the flare process.  

While rapid penumbral decay is a characteristic of the X6.5 flare, it only plays
a very localized role in the M2.0 flare, where slow penumbral decay is the most
prominent feature. Even though Moreton waves have been observed for M-class
flare \citep{Warmuth2004a, Warmuth2004b}, the absence of a wave for the M2.0
flare does not preclude the characterization of X6.5 and M2.0 flares as
homologous. In particular, considering that \citet{Bala2010} associate the
radiant point, i.e., the origin of the Moreton wave more closely with the
satellite sunspot, whereas the centroid of the high-energy flare emission is
more tightly connected to the developing $\delta$-spot. This agree with the
scenario that the M2.0 flare is initiated at the satellite sunspot but the
strong magnetic field gradients in the vicinity of the $\delta$-spot are
responsible for the stronger flare emission.

%###############################################################################
%#
%#    CONCLUSIONS
%#
%###############################################################################

\section{Conclusions}

We have presented a case study involving the flare-prolific active region
NOAA~10930, where a satellite sunspot decayed and flux removal during it was
causally linked to two homologous X6.5 and M2.0 flares. Our major findings with
respect to this slowly decaying sunspot are:

\begin{itemize}
\item[$\square$] Non-radial penumbral filaments indicate the presence of
    twisted magnetic fields in the satellite sunspot.
\item[$\square$] Shear flows were observed along a light-bridge between two
    umbral cores in the center of the satellite sunspot, which is in close
    proximity to the magnetic neutral line. The shear flows continue as long as
    penumbral filaments exist in proximity to the central umbral cores.
\item[$\square$] Slow rotation of the satellite sunspot ($\approx 50\arcdeg$ in
    14~hours), as marked by the tilt angle of the light-bridge, contributes to
    the alteration of the magnetic field topology. 
\item[$\square$] The light-bridge is becoming stronger while the sunspot is
    decaying indicating that it is now easier for convective motions to
    penetrate strong magnetic fields. 
\item[$\square$] Photometric decay rates observed in the satellite sunspot
    are in good agreement with other studies \citep{Bumba1963,
    Moreno-Insertis1988, Pillet1993, Hathaway2008}.
\item[$\square$] We find evidence for localized `rapid' penumbral decay
    \citep{Wang2004, Deng2005, Liu2005} near the central umbral core in
    response to a C6.1 flare. However, `slow' penumbral decay is the more
    prominent characteristic of the decaying satellite sunspot. In particular,
    near northern part of the spot.
\item[$\square$] We find persistent flow kernels with velocities up to
    1~km~s$^{-1}$ close to the region of slow penumbral decay. The decorrelation
    times in this region range from 80-160~min, which are among the longest
    lasting flow structures of the time-series.
\item[$\square$] Even though the intensity-based decorrelation times are high
    for the dominant umbral core (typical values of about 200~min but exceeding
    1000~min in some small kernels), the flow-based decorrelation times are
    significantly lower as compared to the region with slow penumbral decay.
    Therefore, the satellite sunspot decays most noticeably in region with
    weaker and less compact magnetic fields.
\end{itemize}

In summary, we conclude that the decay of the satellite sunspot led to a
substantial restructuring of the magnetic field topology. Thus, flux removal has
to be considered as an important ingredient in triggering flares as we have
discussed in the context of the homologous X6.5 and M2.0 flare. We used the
phrase ``flux removal" because even though flux submergence might be the more
likely scenario, we cannot exclude that flux cancellation is a contributing
factor. Ultimately, only results from local helioseismology can answer this
question \citep[e.g.,][]{Kosovichev2011}. The presence of the $\delta$-spot
provided the environment for even stronger flare emissions. However, rotation,
twist, and rapid proper motions of this $\delta$-spot will become the hallmark
of the flare-prolific active region NOAA~10930 in the following days.

In addition, we adapted and extended the autocorrelation analysis of
\citet{Welsch2012}  to study the lifetime of intensity and flow features. The
novel approach to aggregate decorrelation times into two-dimensional maps will
be a valuable tool to investigate other dynamic processes of the active and
quiet Sun.

%###############################################################################
%#
%#    ACKNOWLEDGEMENTS
%#
%###############################################################################

\begin{acknowledgements}

\noindent \textit{Hinode} is a Japanese mission developed and launched by
ISAS/JAXA, collaborating with NAOJ as a domestic partner, NASA and STFC (UK) as
international partners. Scientific operation of the \textit{Hinode} mission is
conducted by the \textit{Hinode} science team organized at ISAS/JAXA. This team
mainly consists of scientists from institutes in the partner countries. Support
for the post-launch operation is provided by JAXA and NAOJ (Japan), STFC (UK),
NASA, ESA, and NSC (Norway). The GOES Xray flux measurements were made available
by the National Geophysical Data Center. MV expresses her gratitude for the
generous financial support by the German Academic Exchange Service (DAAD) in the
form of a PhD scholarship. CD was supported by grant DE 787/3-1 of the German
Science Foundation (DFG). The authors would like to thank Drs.\ Na Deng and
Rohan E.\ Louis for carefully reading the manuscript and providing ideas, which
significantly enhanced the paper.
\end{acknowledgements}

%###############################################################################
%#
%#    BIBLIOGRAPHY
%#
%###############################################################################

% \bibliographystyle{../../LaTeX/aa}
% \bibliography{../../LaTeX/an-jour,../../LaTeX/meetu}

\end{document}